# Implementing hosting capacity analysis in distribution networks: Practical considerations, advancements and future Directions

Utkarsh Singh, *Senior Member, IEEE*, and Ahmed Al-Durra, *Senior Member, IEEE*

*Abstract*— Hosting capacity analysis is essential for effective integration of distributed energy resources into distribution systems. This paper discusses hosting capacity analysis with emphasis on various aspects affecting the process. This paper addresses key research gaps, aiming to improve the accuracy, scalability, and practicality of hosting capacity estimation. Standardized methodologies are highlighted as a need to ensure consistent hosting capacity estimation across distribution systems. Validation and benchmarking frameworks are emphasized for evaluating various estimation approaches. The potential of data-driven techniques, is also discussed for enhancing hosting capacity analysis. Real-time and dynamic estimation techniques which account for changing system conditions are explored, as well as the integration of hosting capacity analysis with distribution planning and operations. Uncertainty quantification and risk assessment in hosting capacity analysis are identified as crucial areas, advocating for probabilistic and stochastic modelling. This study also emphasizes the importance of considering multiple DER interactions and synergies in hosting capacity analysis, enabling a comprehensive understanding of system performance. This survey aims to serve as a valuable resource for researchers, practitioners, and policymakers, providing insights into advancements made and guiding future research efforts to address identified gaps.

## I. INTRODUCTION

THE rapid growth of distributed energy resources (DERs) in distribution systems pose new challenges for the Distribution System Operators (DSOs). With the increasing presence of solar photovoltaics, wind turbines, and battery energy storage systems (BESS) on the grid, DSOs face the need to accurately assess the hosting capacity of their distribution networks [1]. Hosting capacity analysis (HCA) plays an important role in determining the maximum amount of DERs that can be accommodated without causing detrimental impacts on the stability, reliability and the quality of power supply. Early HCA methods rely on simplified engineering models and deterministic approaches [2]. However, the complex nature of distribution systems, coupled with the uncertainties associated with DER behaviour and system conditions, calls for more advanced and comprehensive methodologies. This has led to significant research interest in developing improved techniques for hosting capacity estimation (HCE). Several research surveys already exist on HCA, focusing on – definitions [3], [4], indices [5], tools [6], [7], enhancement techniques [8], [9], calculation methodologies [10], [11], artificial intelligence applications [12], industry best practices and standardization frameworks [13].

This survey contributes towards the United Nations Sustainable Development Goals (UN-SDGs) 7 (Affordable and Clean Energy), 9 (Industry, Innovation and Infrastructure), and 13 (Climate Action). HCA is essential for integration of distributed energy resources (DERs), such as solar panels and wind turbines in distribution networks. DERs can help to reduce greenhouse gas emissions and mitigate climate change, and they can also support the development of new industries and job opportunities. Motivated by the need to bridge the gap between traditional approaches and the evolving challenges faced by DSOs, this paper aims to serve as a one stop reference for academic and industrial researchers on the rapid advancements and upcoming trends in HCA. The main contributions are:

- **The overview of advancements:** This survey offers a comprehensive overview of the advancements made in HCA in distribution systems. It covers a wide range of research studies and methodologies employed in the field.

- **Research gaps:** This survey identifies key research gaps and limitations in HCA. These gaps will help to highlight potential areas where further research is needed.

- **Future research directions**: This survey aims to outline future research directions and potential areas of improvement in HCE. It identifies emerging trends and technologies that can enhance the accuracy, scalability, and practicality of HCA.

- **Practical challenges in implementation**: This survey also focuses on practical challenges faced by DSOs in HCE. It aims to provide insights into the operational, economic, and technical considerations that should be considered.

- **Informed decision-making**: This survey aims to serve as a valuable resource for researchers, practitioners, and

The authors are with the EECS Department, Khalifa University, Abu Dhabi, UAE [e-mail: utkarsh.singh@ieee.org; ahmed.aldurra@ku.ac.ae, Corresponding Author: Utkarsh Singh]. **This preprint is currently under review and is not yet peer-reviewed. The manuscript may undergo significant changes before final publication. Please cite the published version of the work when it becomes available.**

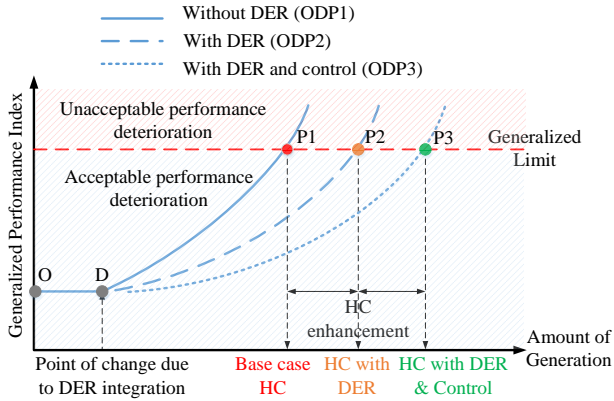

**Fig. 1.** Generalized representation of HC concept. O, D, P1, P2, P3 are operating points on performance curves ODP1, ODP2, and ODP3. O represents the base performance index of any steady-state system parameter. D is the point where DER is integrated and performance index starts deteriorating. P1, P2, and P3 represent the system HC under respective conditions. The performance curves show the the generalized performance index at different DER integration levels. HC can be enhanced by DER integration if there is no violation, and can be enhanced further by advanced control strategies such as voltage control, reconfiguration, and curtailment.

policymakers by offering insights into the state-of-the-art methodologies and findings that can lead to informed decision-making for DER integration and grid planning.

This paper is organized as follows: Section II dives into the overview and fundamentals of HCA. Section III describes the technical advancements in detail. Section IV focuses on the aspects of uncertainty and risk estimation. Section V discusses the scenario of multiple DER interactions, modeling and analysis. Section VI highlights the research gaps and future directions. The paper concludes in Section VII.

## II. HOSTING CAPACITY ANALYSIS: OVERVIEW AND FUNDAMENTALS

The fundamental elements of HCA include assessing the impact of DERs on key distribution system parameters such as voltage profiles, power flows, and equipment limits. This section explores the definitions and key constraints related to HCE.

### A. Definition and Main Aspects

HCA refers to the assessment of the maximum amount of DERs that can be accommodated within a specific section of the distribution grid without compromising its performance, reliability and stability [5]. HC serves as a critical concept in determining the grid's ability to integrate the DERs effectively. A generalized representation of HC concept is shown in Fig. 1. The HC for a system can be represented using performance curves which describe the variation of a given index at different DER integration levels. A generalized performance index may be described as a common minimum index of all the performance parameters in the downstream network. HC is not a fixed calculation with a singular value – but a set of values corresponding to the operational limits of various operational parameters such as voltage and frequency variations, thermal overload, power quality (PQ) and protection problems. Thus, knowing the HC of the system allows the utilities to assess interconnection requirements and identifying areas with low HC enabling upgrades and targeted grid investments. The significance of HCA lies in its ability to support DER integration (i.e. energy transition) while ensuring a reliable grid operation in order to meet future energy demands in a cleaner and more decentralized way.

Obtaining HC maps for a distribution network can help in improving asset utilization, streamlining DER interconnections, and deferring unnecessary upgrades. These are colour coded visual representations of the grid classifying the 3-phase feeder capacities [14] and nodal/ zonal capacities for solar photovoltaic (PV), electric vehicle (EV), and BESS hosting [15]. These maps are mainly based on performance indices and existing capacity information. Protection and reliability aspects are yet to be incorporated. A detailed HCA roadmap with HC map display ideas are provided in [16].

HC can be improved by using advanced control strategies such as distributed control, demand response, energy storage, and market-based mechanisms [17], [18]. For example, distributed control can be used to coordinate the operation of DERs to improve voltage regulation and reduce the stress on the system. Demand response can be used to reduce the load during peak times, which can also help to improve voltage regulation. Energy storage can be used to store energy during off-peak hours and release it during peak hours, which can help to reduce the need for new generation capacity. Market-based mechanisms can be used to allocate resources and optimize the operation of the system, which can also help to improve the system HC.

### B. Key Considerations and Constraints

Theoretically, HCA may be done at node, feeder, substation, area, and system level. Voltage level, loads characteristics and DER type are the primary basis for HC study in distribution networks. Categorical considerations for these are provided in Table 1. These considerations give initial overview of the situation, whereas the actual problem formulation needs understanding case-specific constraints. The number of constraints increase as we move from node to system level, thereby increasing the complexity [9]. Present HCA efforts are mostly focused on node and feeder level. The key constraints which influence HCA can be listed as follows:

- **Operational constraints**: The operational constraints are the most important constraints in HCA, as they ensure the safe and reliable operation of the system. They include voltage limits, thermal limits of equipment, power quality standards, and protection system limitations [19]. These constraints are typically represented as limits on the voltage, current, and power flow in the system. For example, the voltage at each node in the system must be within a prescribed range, and the current flowing through each line must be less than the line's capacity [20].

- **Ancillary services**: Ancillary services are services that are provided by the grid to ensure its reliable operation. These services include frequency regulation, voltage control, and reactive power support [21]. The ability of DERs to provide ancillary services can be considered as a function of the DERs' capacity and controllability [22], [23].

**Table 1.** Key considerations for HCA in distribution networks

| HCA Basis | Category | Considerations | Description |
|---|---|---|---|
| Voltage level | Low voltage (LV) | Voltage fluctuations | LV networks are sensitive to voltage variations, and excessive DERs can lead to voltage problems like over-voltage or under-voltage. |
| | | Load balancing | Balancing loads within LV grids becomes more challenging as DER penetration increases. |
| | Medium voltage (MV) | Substation capacity | HC depends on the capacity of substations and transformers. Upgrades may be required to accommodate more DERs. |
| | | Reliability | MV grids are more resilient to voltage fluctuations, making them a preferred choice for industrial-scale DERs. |
| Load characteristics | Magnitude and variability | Load diversity | Diverse loads, such as a mix of residential, commercial, and industrial consumers, can affect HC differently. |
| | | Load profiles | Analyzing load profiles helps identify periods of high and low demand, which impact HC requirements. |
| | Seasonal load variations | Holiday peaks | HCA may need to account for seasonal spikes in energy demand during holidays or special events. |
| | | Weather related impacts | Extreme weather conditions, such as heatwaves or cold snaps, can lead to significant load variations. |
| DER type | Solar PV systems | Generation patterns | Solar PV generation varies throughout the day, therefore impact of shading and panel orientation on HC must be considered. |
| | | Integration challenges | HCA for solar PV systems involves evaluating their siting, sizing, impact on PQ and grid stability. |
| | Wind turbines | Wind variability | Due to intermittent nature it is important to analyze how this variability impacts HC and grid stability. |
| | | Interconnection and interaction | Siting, sizing, PQ and interaction with other grid-connected DERs must be analyzed. |
| | BESS | Grid support | Battery storage systems can provide grid support services like frequency regulation, enhancing the HC value. |
| | | Charging/ discharging rates | The rate at which batteries charge and discharge affects their impact on HC during peak demand. |
| | EV | Charging infrastructure and patterns | Evaluate the impact of different charging levels and patterns (e.g., Level 1, Level 2, DC fast charging) on HC, as they have varying power demands. Consider the rate of EV adoption in the region and its growth trajectory. |
| | | Vehicle-to-Grid (V2G) potential | Assess the impact of V2G on grid stability, particularly during peak demand or grid emergencies. |

- **DER characteristics**: The characteristics of DERs, including their type, capacity, generation profiles, and variability, have a significant impact on HCA [24]. These constraints are important for ensuring that the DERs can be integrated into the system without causing any problems. These characteristics can be represented as a set of parameters, such as the DERs' power output, the variability of their power output, and their capacity factors [25].

- **Load characteristics**: Factors such as load profiles, diversity, and variability significantly influence the capacity available for accommodating DERs [26]. Load data analysis and modeling techniques help capture the temporal and spatial variations in load demand. These characteristics can be represented in HCA as a set of parameters, such as the load power demand, the diversity of their power demand, and their variability [27].

- **System configuration**: The configuration and topology of the distribution system also influences HC. The spatial distribution of DERs, the location of existing infrastructure (substations, transformers, feeders), and the system layout affect the HCE. Grid modeling techniques, network reconfiguration, and power flow analysis are used to assess the system's capability. These factors can be represented as a set of parameters, such as the number and location of DERs, the number and location of substations, and the length and capacity of the lines [28].

- **Economics**: The existing grid infrastructure and regulatory environment form the basis for HC enhancement. The economics of DER integration are also important to consider in HCA [29], [30]. The cost of DERs, the cost of grid upgrades, and the cost of lost load all need to be considered when making decisions about DER integration.

By considering all of these constraints, it is possible to mathematically model a holistic HCA problem and choose the appropriate tools and techniques for proper estimation.

## III. ADVANCEMENTS IN HOSTING CAPACITY ANALYSIS

HCA is a rapidly evolving field that is being driven by advancements in data analytics, machine learning, optimization algorithms, and computational power. In addition to these, the integration of advanced sensing and monitoring technologies, such as smart meters, phasor measurement units, and distribution automation systems, also facilitates HCA. These technologies provide real-time data on system operating conditions, voltage profiles, power flows, and grid disturbances, enabling dynamic and adaptive HCA. Additionally, the incorporation of grid flexibility options, such as demand response, energy storage systems, and smart grid infrastructure, has expanded the horizons of HCA. By considering the capabilities of these resources, the HC of distribution systems can be further optimized, leading to increased DER integration and improved system performance. In the following sections, these advancements are discussed in detail and their implications for the future of HCA have been explored.

**Table 2.** Standardized methodologies and guidelines

| Resource | Organization | Type of Resource | How it can help in HCA? |
|---|---|---|---|
| Hosting Capacity Guidebook: 2022 Edition [32] | EPRI | Guidebook with technical results | Describes HC process, impact factors (voltage, thermal, protection, PQ and reliability), methods (power flow, direct calculation, model-less, time-based) and guidance with extended examples. |
| Advanced Hosting Capacity Analysis [33] | NREL | Methodology with use cases | Provides HCA methodology for policy makers, utilities and developers mainly in context of distributed PV systems. Discusses static, dynamic uncoordinated and coordinated, i.e. traditional, firm and flexible interconnection HC approaches. |
| Optimizing the grid [34] | IREC | Guidelines | Provides guidelines for optimizing DER integration, efficient decision making, bridging information gaps. Discusses fundamentals, use cases and applicable methodologies (streamlined, iterative, stochastic). HC maps and data sharing are also highlighted. |
| Solution Exchange on Hosting Capacity [35] | DoE | Tools and methodologies | Insights into various tools and considerations related to HCA in the context of bulk power systems. It highlights the challenges and potential solutions for providing accurate and timely data for renewable energy integration. |
| IEC 61968-5: 2020 [36] | IEC | Standard | Defines the enterprise interfaces for distributed energy resource management systems (DERMS). Helps in HCA by providing standardized interfaces, promoting interoperability, facilitating accurate data exchange, and supporting multi-level and multi-party monitoring and control. |
| IEC TS 63276 [37] | IEC | Standard, Guideline, Technical Specification | Upcoming guidelines for HCA. Guideline for the hosting capacity evaluation of distribution networks for distributed generations. It can help to ensure that HCA is conducted in a consistent and transparent manner. |
| IEC 61850/61499 [38] | IEC/ IEEE | Standards based guideline | While the primary focus of this work is on the control and implementation of distributed energy resources, the standardized approaches, reference models, and implementation guidelines it provides can be valuable for HCA. |
| IEEE 1547.9-2022 [39] | IEEE | Standard, guidelines | Provides a foundation of technical standards and guidelines for the safe and reliable interconnection of DERs. This standard is essential in HCA to evaluate the capacity of the grid to accommodate additional DERs while maintaining grid integrity. |

### A. Standardized Methodologies and Guidelines

Standardized methodologies and guidelines must be adopted in HCA to follow industry best practices, regulatory requirements, and technical standards, promoting consistency and comparability. Such guidelines and frameworks describe the data that needs to be collected, the methodologies that need to be used, and the assumptions that need to be made. Although HCA does not have very concrete standards yet, but various organizations such as the Electric Power Research Institute (EPRI) [31], National Renewable Energy Laboratory (NREL) [32], Interstate Renewable Energy Council (IREC) [33], and Department of Energy (DoE) [34], have been actively involved in promoting international collaboration and developing guidelines for HCA. Table 2 provides an overview of currently available or upcoming methodologies and guidelines.

International standards bodies have also indirectly contributed to standardized HCA. For example, IEC 61850, a standard for communication and integration of DERs in smart grids [35], which incorporates hosting capacity assessment guidelines. Another important IEC TS 63276 is also currently under circulation which gives the guidelines for HCA [36]. Similarly, IEEE also provides guidelines for interconnecting DERs with electric power systems, which can be considered for

HCA [37], [38]. The development of standardized methodologies and guidelines for HCA is an important step in the advancement of this field. These methodologies and guidelines provide a common framework for HCA, which can help to improve the accuracy, consistency, and transparency of HCA results. They also help to promote regulatory compliance and stakeholder confidence in HCA.

### B. Validation and Benchmarking Frameworks

Benchmarking is important to ensure consistent and comparable results using different tools and techniques for HCA. An example is shown in [39], where the authors have benchmarked various aspects related to the HC of PV systems using non-synthetic European LV test feeders with actual network and smart meter data. More such efforts are needed with specific focus on HCA, to facilitate informed decision-making for network planners and operators when integrating DERs. Utilities can benchmark their hosting capacity and HCA results to other utilities or to previous years to identify areas for improvement, potential errors, and progress. International collaborations and research projects have also contributed to the development of validation and benchmarking frameworks. For example, the European Union-funded facility DERlab [40], offers several case studies from smart grid and DER integration projects and also provide validation and benchmarking solutions for HCA tools and methodologies. Their frameworks involve real-time hardware-in-the-loop simulations and comprehensive performance evaluation. The usefulness of an HCA is highly contingent on having confidence that its results accurately reflect grid conditions. Utilities can perform validation independently or regulators can oversee these efforts, for example by requiring a utility to submit a validation plan [41]. Table 3 describes a validation framework combining various aspects of HCA

**Table 3.** Validation framework for HCA

| Type of validation | Why validation is needed | Common associated issues | HCA validation procedures |
|---|---|---|---|
| Baseline model | To ensure that the HCA model accurately represents the real-world distribution system. | Errors in input data, such as voltage base, node voltages, loading, equipment default settings, short circuits, circuit reactive power, circuit losses, and aggregate active power. | Check voltage base, voltage at nodes, loading, equipment default settings, short circuits, circuit reactive power, circuit losses, and aggregate active power. |
| Topology | To ensure that the HCA model accurately represents the physical topology of the distribution system. | Errors in GIS data, such as unintentional islands, unintentional meshes, incorrect phase loadings, incorrect feeder switching states, incorrect phases for voltage correction equipment, and the location of existing DERs. | Check for unintentional islands, unintentional meshes, incorrect phase loadings, incorrect feeder switching states, incorrect phases for voltage correction equipment, and the location of existing DERs. |
| Equipment | To ensure that the HCA model accurately represents the performance of equipment in the distribution system. | Errors in equipment data, such as substation data for default settings, substation regulator or load tap changer, line regulators, and capacitors. | Check substation data for default settings, substation regulator or load tap changer, line regulators, and capacitors. |
| Conductor | To ensure that the HCA model accurately represents the performance of conductors in the distribution system. | Errors in conductor data, such as reactive power, circuit losses, voltage drops at peak load, and short circuit currents. | Check reactive power, circuit losses, voltage drops at peak load, and short circuit currents. |
| Consumption & generation profiles | To ensure that the HCA model accurately represents the load and generation profiles of customers on the distribution system. | Errors in customer data, such as reactive power allocation and load allocation. | Check reactive power allocation and load allocation. |
| HCA results | To identify potential errors in HCA results so that they can be corrected before the results are published. | Errors in HCA software, errors in input data, and errors in assumptions. | Check for no (null) or invalid results, zero hosting capacity available, duplicate entries, large discrepancy between previous HCA cycle results and current HCA cycle results, random and spot checks, and additional triggers (e.g., changes in results due to software upgrades or other changes, differences in load profile variation and nodal results that could signal an error). |

## C. Data-driven approaches

Data-driven approaches use historical data to build more accurate models of the distribution network and DERs. This allows them to provide more accurate predictions of the impact of DERs on the network. These techniques utilize large datasets and advanced analytics to address various challenges in assessing and enhancing HC. ML techniques, including regression analysis, support vector machines, and neural networks, are employed to analyze historical data and predict future grid conditions. These models can identify patterns and correlations in load demand and DER profiles that influence HC [42], [43], [44], [45]. Time series analysis is used to understand the temporal variations in grid parameters, such as voltage levels, load and DER profiles [46], [47], [48], [49]. This helps in identifying the optimal times for DER integration. This also forms the basis of most ML applications [42]. Data-centric approaches to HCA can be categorized as model-less or model-based depending on the type of data being used [50]. Fig. 2. shows the flowchart for these approaches and description a comparative assessment of the approaches is discussed here.

**Model-Free Approaches:**

- Grid Measurements: Model-free approaches rely on historical data and real-time measurements from the grid. They don't assume a specific mathematical model of the grid but instead use machine learning algorithms, statistical methods, and data analytics to make predictions about hosting capacity based on observed patterns.

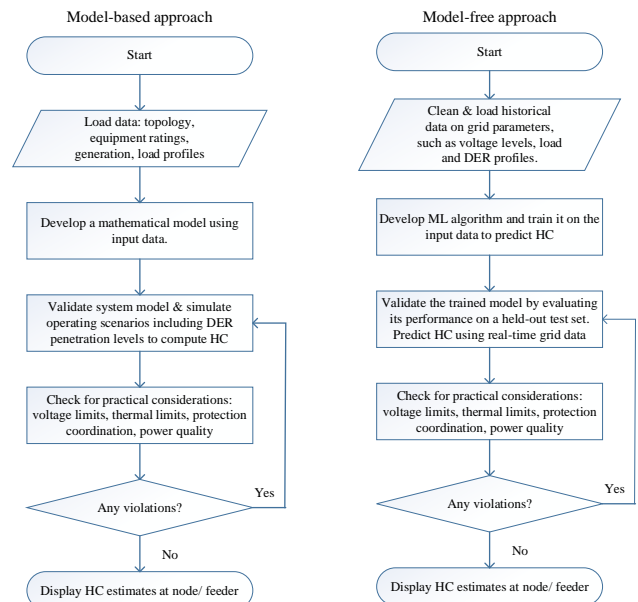

**Fig. 2**. Comparative representations for model-based and model-free approaches in HCA.

- Flexibility: These approaches are more flexible and adaptable to changing conditions and evolving grid configurations. They can capture complex, non-linear relationships that may be challenging to represent with traditional models.

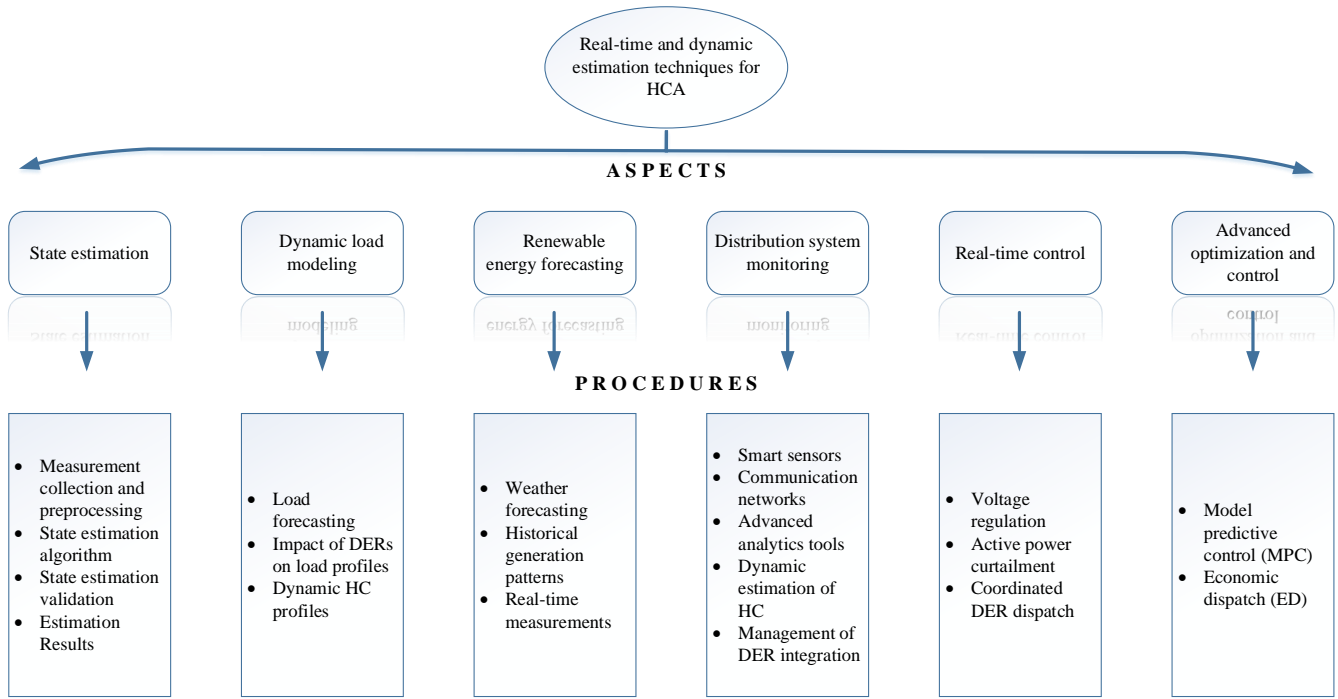

**Fig. 3**. Various aspects related to real-time and dynamic HCA

- Challenges: Model-free approaches often require large amounts of high-quality data for training and may not perform well in situations where data is sparse or noisy. They also lack the transparency of model-based methods, making it challenging to understand the underlying reasons for their predictions.

**Model-Based Approaches:**

- Mathematical Models: Model-based approaches rely on mathematical models of the power distribution network. These models represent the physical and operational characteristics of the grid, including line impedance, transformer ratings, voltage limits, and more.

- Accuracy and Transparency: Model-based approaches are typically more accurate when the underlying model represents the real-world system accurately. They provide transparency because results are derived from known equations and parameters.

- Challenges: These approaches may struggle when the grid undergoes significant changes or when they encounter uncertainties not explicitly included in the model. Maintaining up-to-date models can also be a challenge.

### D. Real-time and dynamic estimation techniques

Fig. 3. shows the various aspects related to ensuring real-time and dynamic HCA. One of the key techniques used is state estimation. State estimation utilizes measurements from system sensors and SCADA (Supervisory Control and Data Acquisition) data to estimate the current state of the distribution system, including voltage magnitudes, angles, and power flows [51]. By incorporating real-time measurements, state estimation techniques can provide accurate information for HCA, allowing the identification of system constraints and the determination of available capacity for DER integration. Dynamic load modeling is an important aspect of real-time and dynamic estimation. It captures time-varying behavior of loads and the impact of DERs on the distribution system. These models consider load fluctuations, demand response, and DER output variations, providing dynamic HC profiles under changing system conditions [52, see Fig. 17]. Renewable energy forecasting is crucial for real-time and dynamic HCE. Accurate predictions of solar and wind power generation enable proactive management of DER integration and grid stability [53]. Various ML techniques can be utilized for renewable energy forecasting, considering weather data, historical generation patterns, and real-time measurements. Distribution system monitoring and control are essential for real-time HCA. Advanced metering infrastructure, smart sensors, and communication networks enable the collection of real-time data from DERs, loads, and distribution equipment. This data, combined with advanced analytics, can facilitate dynamic estimation of HC, taking into account system conditions, DER behavior, and grid constraints [54]. Real-time control strategies, such as voltage regulation, active power curtailment, and coordinated DER dispatch, can be employed to manage DER integration within the HC limits. Furthermore, real-time and dynamic estimation techniques benefit from advanced optimization and control algorithms. Model Predictive Control (MPC) and Economic Dispatch (ED) methods optimize DER operation and scheduling in real-time, considering system constraints and objectives [55]. These techniques enable the efficient utilization of DERs while maintaining system reliability and stability.

### E. Integration with Distribution Planning and Operations

By integrating HCA, distribution planners can identify the suitable locations and sizes for DER installations, considering the maximum integration potential and system constraints. Furthermore, integration with distribution operations allows for real-time monitoring and management of DERs within their HC limits [33]. DSOs can utilize real-time data, advanced analytics, and control strategies to optimize DER dispatch, ensure voltage and frequency regulation, and manage the system's stability and reliability [14], [15]. Integrated distribution management systems (IDMS) can play a crucial role in the integration of HCA with distribution planning and operations. IDMS platforms bring together advanced data analytics, optimization algorithms, and control functionalities to facilitate decision-making and real-time control of DER integration [56]. These systems can enable DSOs to monitor DER performance, estimate HC in real-time, and implement proactive control strategies to manage DER integration within the system's limits. The integration of HCA with distribution planning and operations requires the development of appropriate information exchange frameworks and data architectures. A figurative representation is shown in Fig. 4. Standardized data models, communication protocols, and interoperability standards enable the seamless flow of data between HCA tools, distribution planning software, and operational control systems. These frameworks facilitate the integration of HCA into the existing distribution system infrastructure. Additionally, the integration of HCA with advanced grid management techniques, such as demand response programs, distribution automation, and energy storage systems, enhances the flexibility and resilience of distribution systems. Demand response programs enable the active participation of consumers in adjusting their electricity consumption patterns based on HCA constraints. Energy storage systems, coupled with HCA, allow for the optimal utilization of DERs and grid support during peak demand periods [57].

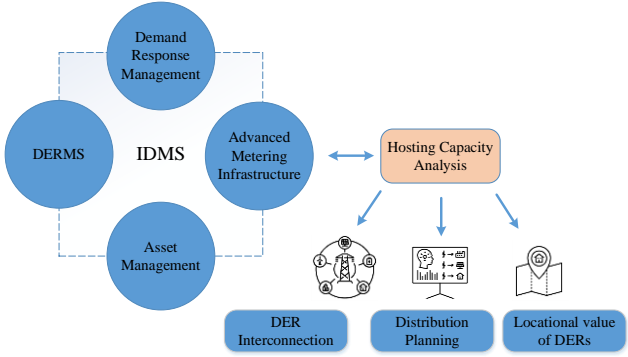

**Fig. 4.** Integrating HCA with distribution planning and operations

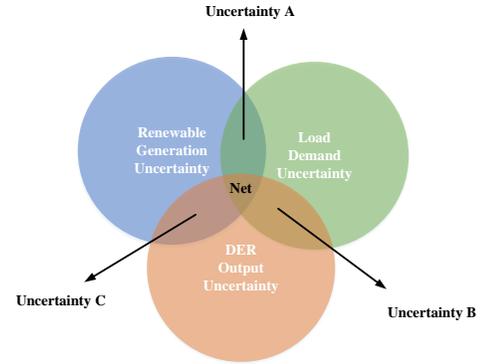

**Fig. 5.** A conjecture of uncertainties and their impact on HCA. 'Uncertainty A' could be a case of sudden decrease in renewable energy generation due to a cloud passing over a solar power plant, the net load on the distribution system will increase. This can lead to problems such as voltage instability and overloaded feeders. 'Uncertainty B' might represent a sudden increase in load demand, the distribution system operator may need to curtail DER output in order to avoid overloading the system. 'Uncertainty C' could be a situation of sudden decrease in renewable energy generation and a simultaneous increase in DER output, the distribution system operator may need to curtail DER output in order to avoid overloading the system. The 'Net' uncertainty is the overall uncertainty that will be reflected in HCE. To account for this, appropriate uncertainty quantification methods must be used in HCA.

## IV. Addressing Uncertainty and Risk in Hosting Capacity Analysis

### A. Uncertainty and Quantification Methods

Uncertainty quantification is a crucial aspect of HCA in distribution systems, as it helps account for the inherent variability and unpredictability of various factors, such as renewable energy generation, load demand, and DER output as shown in Fig. 5. Uncertainty quantification methods employed in HCA and their application in improving the accuracy and reliability of assessments are discussed here:

- **Probabilistic Methods:** Probabilistic methods involve modeling uncertainties using probability distributions and statistical techniques. For instance, Monte Carlo simulation generates random samples of uncertain parameters based on probability distributions (e.g., Gaussian, uniform, or truncated distributions). This approach provides insights into the range of HC parameters and their associated probabilities [58]. Other probabilistic methods, like Latin Hypercube Sampling and importance sampling, can also efficiently explore the uncertainty space.

- **Sensitivity Analysis:** Sensitivity analysis assesses the influence of uncertain input parameters on HC estimates. It can help in identifying the most influential factors affecting HC and offers valuable insights for system planning and operation [59]. Global sensitivity indices like Sobol' indices quantify the contribution of each input parameter to the overall output variability, aiding in prioritizing uncertainties for further analysis [60]. Local sensitivity analysis techniques can help in examining the sensitivity of HC to small variations in input parameters.

- **Interval Analysis:** Interval analysis quantifies uncertainties by defining input parameters within intervals instead of specific probability distributions. This method considers worst-case scenarios within the specified intervals, which can give more conservative HC estimates [61]. Interval arithmetic and interval optimization techniques compute upper and lower bounds of hosting capacity, offering a robust range of values resilient to uncertainties.

- **Fuzzy Logic:** Fuzzy set theory represents uncertainties in linguistic terms, facilitating the modeling of subjective and imprecise information. This approach has also been presented as possibilistic approach in [62]. Fuzzy logic-based HCA incorporates fuzzy parameters like membership functions, fuzzy rules, and fuzzy inference systems to capture the vagueness and ambiguity of uncertain inputs.

**Table 4**. Risk assessment frameworks in HCA

| Framework | Description | Why is it needed? | Common sources of risk | Key methods |
|---|---|---|---|---|
| Reliability analysis | Assesses the ability of the distribution system to meet reliability criteria under various operating conditions, considering the impact of DERs. | To ensure that the distribution system can meet the needs of its customers, even when there are disruptions to service. | Load growth, equipment failures, extreme weather events, and cyberattacks. | Probabilistic reliability analysis, component reliability analysis, resilience analysis, and risk-based reliability analysis. |
| Voltage and stability analysis | Assesses the impact of DERs on voltage and stability of the distribution system. | To ensure that the distribution system operates within safe and stable limits. | Changes in load demand, DER output, and system configuration. | Voltage analysis, voltage stability analysis, harmonic analysis, and transient stability analysis. |
| Thermal analysis | Assesses the impact of DERs on the thermal capacity of the distribution system. | To prevent equipment overheating and thermal constraints violation. | Increased load demand, DER output, and ambient temperature. | Load flow-based thermal analysis and dynamic thermal analysis. |

- **Bayesian Inference:** Bayesian inference methods integrate prior knowledge and observed data to update probability distributions and estimate posterior distributions of uncertain parameters. Bayesian approaches provide a systematic framework to incorporate expert knowledge and historical data into HCA [63]. These methods enhance the accuracy of HC estimates by effectively integrating available information and data.

Through these methods, a deeper understanding of the impact of uncertainties can be gained, leading to reliable HC estimates.

### B. Risk Assessment Frameworks

**1. Reliability Analysis:** Reliability analysis is a fundamental aspect of HCA, ensuring the secure and dependable integration of Distributed Energy Resources (DERs). This section outlines key reliability analysis methods within HCA.

- **Probabilistic Reliability Analysis:** It can help in assessing the probability of meeting system reliability criteria considering uncertainties in load demand, DER output, and system configuration. Monte Carlo simulation can be used to generate random samples and simulate system behavior, aiding in identifying vulnerabilities and optimizing system configuration [64].
- **Component Reliability Analysis:** This focuses on assessing the reliability of individual components within the distribution system, incorporating failure rates and dependencies [65]. Techniques like fault tree analysis and Bayesian networks can be used to quantify component reliability and improve overall system reliability.
- **Resilience Analysis:** It evaluates the system's ability to withstand and recover from disruptions, such as impact of driver behavior on EV [66]. For active DER participation, mathematical modelling of incentives and penalties can be included. Stochastic optimization and game theory can be used to identify optimal configurations and response strategies in such type of analysis.
- **Risk-Based Reliability Analysis:** Combining probabilistic and risk assessment approaches, this method considers the likelihood and consequences of various events, aiding in risk quantification and resource allocation. For example, multi-period reconfiguration planning considering HC provision [67].

**2. Voltage and Stability Analysis:** Voltage and stability analysis is crucial to maintain voltage quality and system stability during HCA.

- **Voltage Analysis:** Load flow analysis, sensitivity analysis, and optimal power flow and voltage profile evaluation must be done to keep voltage within acceptable limits [68].
- **Voltage Stability Analysis:** This aspect examines the system's ability to maintain voltage stability during normal and contingency conditions, utilizing methods such as voltage stability margin analysis [69].
- **Harmonic Analysis:** This deals with the impact of harmonics on HCA, assessing levels and potential impacts on system performance [70].
- **Transient Stability Analysis:** Addition of dynamic loads reduces the HC when based on minimum voltage swing limits. Thus, it is necessary to evaluate the impact of transient stability effects on HC [71].

**3. Thermal Analysis:** Thermal analysis ensures that the integration of DERs doesn't lead to equipment overheating and thermal constraints violation.

- **Load Flow-Based Thermal Analysis:** Assesses steady-state thermal conditions using load flow analysis integrated with thermal models, considering equipment ratings and thermal constraints [72].
- **Dynamic Thermal Analysis:** This refers to HC analysis and enhancement by using dynamic thermal rating of an equipment [73].

Incorporating these risk assessment frameworks in HCA can enable DSOs to better understand and mitigate potential risks associated with integrating DERs into their systems. An overview of these frameworks in presented in Table 4.

### V. MULTIPLE DER INTERACTIONS IN HOSTING CAPACITY ANALYSIS

#### A. Synergistic Effects and Challenges

The combination of DERs with complementary generation profiles, such as solar PV and wind turbines, can increase the overall energy production and reduce the reliance on the grid. This synergy allows for a higher HC by maximizing the utilization of available renewable energy resources. The integration of energy storage systems with DERs can mitigate the intermittent nature of renewable generation and provide additional flexibility to the system [74]. By storing excess energy during periods of high generation and discharging it during peak demand, the hosting capacity of DERs can be increased while ensuring grid stability. Coordinating DERs with demand response programs enables load shifting and demand management. By adjusting the consumption patterns of end-users based on the availability of DERs, the HC can be optimized, and grid congestion can be alleviated [75]. Thus, interactions among multiple DERs require careful coordination and control mechanisms. The challenge lies in developing efficient algorithms and communication protocols to optimize their collective operation while maintaining system stability and reliability. The intermittent nature of renewable energy sources and the varying demand patterns pose challenges in accurately predicting and managing the synergistic effects. Advanced analysis and real-time monitoring systems are essential to address these uncertainties. Incorporating synergistic effects in distribution system planning and infrastructure development is a challenge. System operators and planners need to consider the optimal location, sizing, and interconnection of DERs to maximize their combined benefits and avoid potential conflicts. The existing regulatory and market frameworks may not adequately capture the value and potential of synergistic effects [76]. Therefore, new policies and market mechanisms are necessary to incentivize and support the integration of multiple DERs and promote their combined benefits in HCA.

**Table 5**. An overview of the methods used in HCA

| Method | When to choose this method | Advantages | Disadvantages | Approach to HCA |
|---|---|---|---|---|
| Analytic method | When detailed and accurate results are required, and the time and resources are available. | - Provides accurate and detailed results. <br> - Can be used to evaluate a wide range of scenarios and constraints. | - Can be computationally expensive. <br> - Requires detailed grid data and expertise in power system modeling and analysis. | 1. Create a detailed model of the distribution system, considering all relevant components and their control strategies. <br> 2. Identify the key constraints that affect HC, such as voltage limits, thermal limits, and transient stability. <br> 3. Use the chosen method and simulate the system operation under various conditions. <br> 4. Analyze the simulation results to calculate the HC and identify potential constraints. |
| Stochastic method | When it is important to account for uncertainties in DER output and load profiles. | - Can account for uncertainties in DER output and load profiles. <br> - Provides probabilistic estimates of HC, which can be useful for decision-making. | - Can be computationally expensive. <br> - Requires well-defined probability distributions for DER output and load profiles. | |
| Streamlined method | When a rapid assessment of HC is needed, or when data and resources are limited. | - Fast and easy to implement. <br> - Can be used with limited data and resources. | - May not be as accurate as the analytic or stochastic methods. <br> - May not be suitable for complex systems or when a high level of accuracy is required. | |

*B. Modeling and Analysis Techniques*

HCA involves a holistic approach to modeling and assessing distribution systems' ability to accommodate DERs. It begins with accurate distribution system modeling, considering components like transformers and cables, followed by power flow analysis, dynamic stability assessment, and harmonic analysis to ensure system reliability and power quality compliance. With proper system modeling and key constraints considered, following methods can be used for HCE [1]-[7]:

- **Analytic Method**: This method typically involves detailed power system modeling and simulation using power flow and dynamic analysis tools. It is a deterministic approach that calculates hosting capacity by analyzing the steady-state and dynamic behavior of the grid under various conditions. Analytic methods consider factors like voltage limits, thermal limits, and transient stability. This method provides precise results but may require extensive computational resources and detailed grid data.
- **Stochastic Method**: These methods use probabilistic elements in HCA. Monte Carlo simulations and probabilistic power flow analysis are common techniques used within the stochastic method. Stochastic methods account for uncertainties in DER output and load profiles, providing probabilistic estimates of HC. They are valuable for assessing HC under variable and uncertain conditions.
- **Streamlined Method**: This method often uses simplified models and heuristics to assess HC without the need for extensive grid modeling or detailed simulations. Streamlined methods are valuable for initial screening and feasibility studies, as they offer a more rapid assessment of hosting capacity. They may not capture all nuances of the grid behavior but are useful for identifying potential HC constraints.

While these three methods are the most commonly discussed in the literature, it's important to note that there is no one-size-fits-all approach to HCA. The choice of method depends on the specific goals, data availability, and the level of detail required for a particular analysis. Some analyses may use a combination of these methods based on the guidance provided in Table 5, to provide a comprehensive understanding of HC within a given grid area.

## VI. RESEARCH GAPS AND FUTURE DIRECTIONS

HCA is an evolving field, and several emerging trends and potential areas of improvement can further enhance its effectiveness and applicability. This section highlights the key areas or gaps identified, which need further efforts, along with specific inputs for future research:

- **Time-variant and dynamic conditions:** Dynamic HCA techniques should be developed to consider real-time variations in load, generation, and system conditions. These techniques should integrate advanced forecasting algorithms and real-time monitoring data to accurately capture temporal dynamics.

- **Distribution system constraints:** The scope should be broadened in HCA to encompass not only voltage limits but also thermal limits, line losses, and voltage stability. Furthermore, advanced optimization algorithms should be developed to accommodate multiple operational constraints simultaneously.

- **Uncertainty and risk quantification:** To quantify uncertainty and risk effectively, researchers should investigate methods such as probabilistic modeling and scenario-based analysis. Additionally, they should develop frameworks for assessing risks to evaluate the impact of DER uncertainties on system reliability.

- **Integration of advanced ML Techniques:** Researchers should incorporate advanced deep learning, reinforcement learning, and knowledge-based recommender systems to handle multi-dimensional datasets effectively and provide context-aware HC estimates.

- **Coordinated planning and operation**: Efforts should focus on exploring methods for coordinating DERs with distribution planning and operation. This entails considering optimal DER placement, assessing grid support capabilities, and evaluating flexibility services.

- **Standardization and interoperability**: It is imperative to foster collaboration among stakeholders to establish common standards and frameworks. Additionally, efforts should encourage the development of open-access datasets and interoperable software tools.

- **Validation and benchmarking frameworks**: Robust validation and benchmarking frameworks for HCA techniques need to be established. These frameworks will enable accurate and reliable assessment and comparison of different methods.

- **Data analytics and validation**: Researchers should develop advanced data analytics and validation techniques, including data fusion, data mining, and anomaly detection, for enhanced data utilization and reliable HC estimates.

- **Accounting for multi-DER interactions**: Comprehensive research should be conducted to analyze the synergistic effects and challenges arising from the integration of multiple DERs and how it influences dynamic HC.

- **Collaborative planning and operation:** Promotion of information exchange, data sharing, and collaborative decision-making processes among stakeholders is essential for more accurate HC assessments.

- **Resilient and flexible infrastructure**: Efforts should be directed toward assessing the flexibility potential of DERs, including demand response, energy storage, and vehicle-to-grid integration. This will contribute to the design and deployment of resilient and flexible distribution infrastructure.

- **Enhanced monitoring and control**: Integration of real-time data from advanced sensors and smart grid infrastructure into HCA should be prioritized to provide more accurate and up-to-date results for efficient DER integration.

- **Integration of market mechanisms:** Market mechanisms like locational marginal pricing and demand-side participation should be incorporated into HCA to optimize the economic dispatch of DERs and their integration into wholesale electricity markets.

- **Integration of communication and control technologies**: Exploration of the integration of technologies like IoT, smart grid, and advanced metering infrastructure (AMI) is essential to enhance monitoring and control. This optimization of the HCA process will enable proactive system management.

- **Policy and regulatory frameworks**: Supportive policy and regulatory frameworks need to be developed to incentivize DER deployment, facilitate market participation, and promote grid modernization. This acceleration will drive the adoption of HCA and transition to more sustainable distribution systems.

- **Hybrid modeling approaches**: Combining physics-based models with data-driven approaches is crucial for comprehensive and accurate HCA. Leveraging the strengths of both approaches enables integrated DER analysis.

- **Field validation and case studies**: Extensive field validation studies and case studies in real-world distribution systems should be conducted to validate HCA methods. These practical insights help identify challenges, limitations, and potential improvements in deployment scenarios.

## VII. CONCLUSION

This paper presented an overview of the practical considerations essential for implementing HCA in distribution systems. By addressing key research gaps and highlighting the need for standardized methodologies, validation frameworks, and data-driven techniques, a roadmap has been outlined for improving the accuracy, scalability, and practicality of HCE. The integration of real-time and dynamic techniques, coupled with the incorporation of HCA into distribution planning and operations, can offer a promising path forward. Additionally, recognizing the significance of uncertainty quantification, risk assessment, and the complexities of multiple DER interactions underscores the importance of probabilistic and stochastic modeling in HCA. With the continuous developments in HCA, this survey provides some key insights towards informed decisions and innovative solutions in the integration of DERs into distribution systems. The future research should focus more on managing uncertainties, developing advanced analysis techniques, and promoting standardization and collaboration to ensure reliable HCA.

## ACKNOWLEDGMENT

This work was supported by a research fund from Khalifa University.

**Utkarsh Singh** (Senior Member, IEEE) received the Ph.D. degree in electrical engineering from the Indian Institute of Technology, Roorkee, India, in February 2018. He is currently a visiting researcher in the Electrical Engineering & Computer Science Department at Khalifa University, UAE. His research interests include power systems, power quality, signal processing, and applied artificial intelligence.

**Ahmed Al-Durra** (Senior Member, IEEE) received the B.Sc., M.Sc., and PhD degrees in Electrical & Computer Engineering from Ohio State University in 2005, 2007, and 2010, respectively. He is a Professor in the Electrical Engineering & Computer Science Department at Khalifa University, UAE. He is also the Associate Provost for Research in Khalifa University, UAE. His research interests are applications of control and estimation theory on power systems stability, micro and smart grids, renewable energy systems and integration, and process control. He has over 260 scientific articles in top-tier journals and refereed international conference proceedings. He has supervised/co-supervised over 30 PhD/Master students. He is leading the Energy Systems Control & Optimization Lab under the Advanced Power & Energy Center, an Editor for IEEE Transactions on Sustainable Energy and IEEE Power Engineering Letters, and Associate Editor for IEEE Transactions on Industry Applications, IET Renewable Power Generation, and Frontiers in Energy Research.